\documentclass[useAMS,usenatbib]{mn2e}
\usepackage{latexsym,graphicx,natbib}
\usepackage{psfig}
\def\aplt{{\th \rlap{\raise 0.5ex\hbox{$\scriptstyle  {<}$}}
    {\lower 0.3ex\hbox{$\scriptstyle  {\sim}$}} \th }}  
\def\apgt{{\th \rlap{\raise 0.5ex\hbox{$\scriptstyle  {>}$}}
    {\lower 0.3ex\hbox{$\scriptstyle  {\sim}$}} \th }}  
\def\th{\thinspace}
\def\ts{{\raise 0.3ex\hbox{$\scriptstyle {\th \sim \th }$}}}

\newcommand{\msun}{{\rm\,M_\odot}}
\def\ergs{{\rm\,erg\,s^{-1}}}
\def\gcm3{{\rm g\,\, cm^{-3}}}

\loadboldmathitalic
\title[The ultra luminous X-ray souce in M82]{The ultra-luminous X-ray source in M82: an intermediate mass black hole with a giant companion}

\author[A. Patruno, S. Portegies Zwart, J. Dewi, C. Hopman] {A. Patruno$^{1\ast}$, S. Portegies Zwart$^{1,2}$, J. \
Dewi$^3$, C. Hopman$^4$\\
\\
\normalsize{$^{1}$Astronomical Institute ``A. Pannekoek'', University of Amsterdam, Kruislaan 403, 1098 SJ, The Netherlands}
\\
\normalsize{$^{2}$Section Computational Science, University of Amsterdam, Kruislaan 403, 1098 SJ, The Netherlands}
\\
\normalsize{$^3$ Institute of Astronomy, University of Cambridge, Madingley Road, Cambridge CB3 0HA}
\\
\normalsize{$^4$ Faculty of Physics, Weizmann Institute of Science, P.O. Box 26, Rehovot 76100, Israel}
\\
\normalsize{$^\ast$To whom correspondence should be addressed; E-mail:
apatruno@science.uva.nl}
}

\date{}

\begin{document}

\maketitle

\begin{abstract}

The starburst galaxy M82 at a distance of 12 million
light years, is the host of an unusually bright $2.4-16\times
10^{40}$\,erg/s X-ray point source, which
is best explained by an accreting black hole $10^{2}$ to $10^4$ times
more massive than the Sun.  Though the strongest
candidate for a so called intermediate mass black hole, the only
support stems from the observed luminosity and the 0.05-0.1\,Hz quasi
periodicity in its signal.  Interestingly,
the $7-12\,$Myr old star cluster MGG-11 which
has been associated with the X-ray source is
sufficiently dense that an intermediate mass black hole could have
been produced in the cluster core via collision runaway.
The recently discovered $62.0\pm 2.5$ day
periodicity in the X-ray source X-1 further
supports the hypothesis that this source is powered by a black hole
several hundred times more massive than the Sun.  We perform detailed
binary evolution simulations with an accreting compact object of
$10-5000\,\msun\,$ and find that the X-ray luminosity, the age of
the cluster, the observed quasi periodic oscillations and the now
observed orbital period are explained best by a black hole of
$200-5000\,\msun\,$ that accretes material from
a $22-25\msun$ giant companion in a state of Roche-lobe contact.
Interestingly such a companion star is consistent with the expectation
based on the tidal capture in a young and dense star cluster like
MGG-11, making the picture self consistent.

\end{abstract}

\section{Introduction}
Ultra Luminous X-ray sources (ULXs) are extra galactic off nuclear X-ray
point sources with an observed X-ray luminosity between $10^{39}$ and
$10^{41}\ergs$, well in excess of the Eddington limit for a stellar
mass black hole ($L_{Edd}=1.25\times 10^{38}(M_{bh}/M_{\odot})\ergs$, with $M_{bh}$ the mass of the accretor). While there is evidence of a binary nature of these
sources, e.g. in M51, in IC342 and in Circinus \cite{2004IAUS..222...21W,2005ApJ...620L..31K},
 the nature of their compact
accretors remains somewhat controversial.
A possibility is represented by beamed models with stellar mass black
holes where a mechanical collimation in a thick disc or a Doppler
boosting from a jet in a microblazar might produce an apparent super
Eddington emission up to a factor $\sim 10$ explaining the nature of
some ULXs with luminosities below $\sim 10^{40}\ergs$ \cite{2001ApJ...552L.109K,
2002A&A...382L..13K,2004PThPS.155...27M}.  Another possibility
can be a genuine super Eddington emission from a stellar mass black
hole with a radiation pressure dominated disc that can reach
luminosities as high as $10^{40}\ergs$ \cite{2002ApJ...568L..97B}.
 However all these models cannot easily explain those sources with luminosities
above $\sim 4\times10^{40}\ergs$ \cite{2005MNRAS.357..275K}.  The presence of an Intermediate Mass
Black Hole (IMBH) with a mass around $10^{2}-10^{4}\msun$ represents an
exciting possibility to naturally explain at least the most luminous
sources. However its existence has not yet been demonstrated. \\
One of the best ULX to search for an IMBH is the persistent
source X-1 in the starburst galaxy M82 \cite{Bode1777} 
whose X-ray brightness can reach values as high as
$2.4-16\times 10^{40}\ergs$ \cite{1999PASJ...51..321M} that
corresponds to bolometric fluxes larger than $5-20\times
10^{40}\ergs$, unexplainable with a stellar mass black hole \cite{2001ApJ...547L..25M,2001MNRAS.321L..29K,2005MNRAS.357..275K}. The detection of a millihertz quasi periodic oscillation
at a frequency of 54 mHz \cite{2003ApJ...586L..61S} supports the
hypothesis of an IMBH as the compact accretor of M82 X-1 \cite{2005MNRAS.tmp.1126M,2005astro.ph..9646D}. A recent {\it RXTE} detection of
an X-ray modulation with a period of $62.0\pm 2.5\,$ days has been
interpreted as the orbital period of the ULX with a donor star on the
giant/supergiant phase with a mean density $\rho\simeq 5\times
10^{-5}\rm\,g\,cm^{-3}$ \cite{2006Scie...tmp}.  Remarkably, the position 
of M82 X-1 is coincident with that of MGG-11, a Young Dense Cluster (YoDeC)
with a mass of $\sim
3.5\times10^{6}\msun$ and an age of $7-12$Myr \cite{2003ApJ...596..240M}.

In recent $N$-body simulations of MGG-11 \cite{2004Natur.428..724P}
it has been shown that in its first 3 Myr, the cluster undergoes a runaway
collisional merger of massive main sequence stars leading to the formation of a
giant protostellar object of several 1000$\msun$ in its core.
The subsequent evolution of this
supermassive star is still poorly known and probably can lead to the
 formation of an IMBH with a mass between several hundred and a
 few thousands solar masses.
Once the IMBH is formed it might be fed by the mass transfer from a donor star
captured via tidal interaction(Hopman, Portegies Zwart \& Alexander 2004, Baumgardt et al. 2005)
 that fills its Roche lobe during the main sequence or on a later stage of the evolution.
\\
If the apparent coincidence of M82 X-1 with MGG-11 connotes that the source X-1 is located in MGG-11, this coincidence can be 
considered to be the key-point to solve the controversy about the origin
of the compact accretor \cite{2004Natur.428..724P}. 
In the here proposed scenario, an estimate of the age is
crucial in determining the mass of the possible donor star in an X-ray
binary with a black hole as compact object. This drastically limits
the various possibilities, and allows us to perform detailed
pin-pointed simulations to constrain the other observables with our
binary evolution simulations.
In Section 2 we describe the evolution and the initial conditions used for our binary simulations. In Section 3 we present our results and discuss them in Section 4.

\section{Initial conditions and evolution of the binary}
At an age of $t \sim 7$\,Myr all the stars with an initial mass
$M>26\msun$ have experienced a supernova explosion and can therefore
be excluded as possible donors to a black hole.  Given the upper age limit of
12 Myr, the low donor mean density
observed implies that the companion star has already left the main sequence,  
limiting the mass to be $M >17\,\msun$. \\

In this paper we aim to investigate the binary nature of this ULX using very
stringent limits on the parameters of the binary:  the X-ray luminosity of M82 X-1 ($2.4-16\times 10^{40}\ergs$), the orbital period of M82 X-1 ($62.0\pm 2.5$days), the age of MGG-11 (7-12 Myr), the mean density of the donor of M82 X-1 ($\simeq 5\times 10^{-5}\rm\,g\,cm^{-3}$).
Assuming that the system has been formed via tidal interaction we can add a
few more constraints on the binary since the mass transfer cannot start before the time required to form an IMBH  ($t\sim 3$Myr, Portegies Zwart \& Mc Millan 2002).
The tidal capture process itself also
puts an interesting constraint to the initial conditions, as capture
can only be successful in a rather narrow range of impact parameters (Hopman, Portegies Zwart \& Alexander 2004).  Subsequent tidal dissipation in the tidally
captured star causes the orbit to circularize to an orbital separation
in the range of $2r_{t}<a_i<5r_{t}$ \cite{2005astro.ph.11752B}, with $r_{t}=(M_{bh}/M)^{1/3}R$ the tidal radius of the black hole, where $M$ and $R$ are the mass and the radius of the donor star.
Larger orbital separations are hardly realizable as those can only
result from an encounter with a larger impact parameter, and those are unable to dissipate sufficient
energy in the tidal interaction \cite{1977ApJ...213..183P}. On the other
hand, an encounter which brings the star too close to the IMBH will
result in the destruction of the incoming star, leaving only a disc of
stellar debris \cite{1988Natur.333..523R,2003ApJ...590L..25A}.
After the tidal capture, the binary continues to evolve on nuclear timescales 
of the captured star and by the emission of gravitational radiation. 
 We adopted an updated version of Eggleton's binary evolution code \cite{2002ApJ...575..461E,1995MNRAS.274..964P} to perform a large number of
simulations in which we vary the mass of the stellar donor (i.e. the
captured star), the mass of the intermediate mass black hole and the
initial orbital separation. We assume that mass transfer on
the IMBH is Eddington limited and that the excess mass is lost from
the system with the specific angular momentum of the
accretor.
For mass loss by the stellar wind we adopt a de Jager like wind \cite{1988A&AS...72..259D}, including the modified
recipe for the wind accretion onto the IMBH during the detached phase \cite{2005MNRAS.364..344P}. We assume that all the stars have Population I chemical composition, mixing length parameter $\alpha = 2.0$ and overshooting constant $\delta_{ov}=0.12$.
We then explored black hole masses between 10 and 5000$\msun$, which
includes the possibility that the accreting object is a stellar mass
black hole. The donor mass is varied between 18 and 26$\,\msun$.
The IMBHs were assumed to acquire their stellar remnant still on the main sequence at an age of $\sim 4\,$Myr in a circular binary with a semi-major axis $a\simeq\,0.2-0.5$AU, according to the tidal capture scenario.
For the stellar mass black hole we used an initial value of the semi-major axis $\aplt 0.2$AU, typical of normal black hole binaries, to start the RLOF phase after $3-4$Myr when the star is still on the main sequence or just at the beginning of the giant branch.

\section{Results}
\begin{center}\begin{figure}
\centerline{\psfig{file=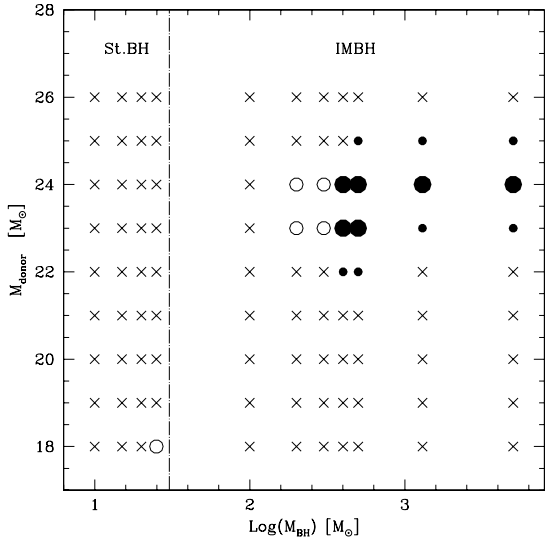,height=3in}} \caption{Grid of binaries used for the simulations.
  The dots are those systems with total agreement with the observations. The close dots are those binaries that
  match with the luminosity of M82 X-1 without any super Eddington or beamed emission. The empty circles are those\
  binaries where a good agreement is reached with the need of beaming or super Eddington emission. When the para\
  meters of the binary are within 1$\sigma$ form the observed period the dots are bigger, while the small one are\
  those in agreement within 2$\sigma$. The crosses are binaries with a discrepancy larger than 2$\sigma$ from th\
  e observed values.} \label{fig.1}
\end{figure} \end{center}

In fig.\,\ref{fig.1} we present the result of these
simulations. The extend to which the simulated binaries comply with
the observed X-ray source is coded in the various symbols. The close
circles in fig.\,\ref{fig.1} identify those initial conditions for our
binary evolution simulations which satisfactorily reproduce the
observations (period and luminosity).

The crosses fail to reproduce the observations.  
The open circles give comparable orbital separation in the observed age
interval of 7 to 12\,Myr, but require super Eddington accretion on the
black hole to comply with the observed X-ray luminosity. This happens
for black holes of $M_{\rm bh} < 400\,\msun$.
The simulation with a $18\msun$ donor and a $\sim25\msun$ black hole
(indicated with the open circle in the lower left corner of
fig.\,\ref{fig.1}) is able to reach the observed orbital period and mass
accretion rate at an age of 11.5\,Myr. This binary started the Roche-lobe overflow (RLOF) at the
age of $\sim 4$\,Myr with the donor still on the main sequence. 
However, in order to comply with the observed
X-ray luminosity the emission has to be beamed with an angular
diameter between 4 and 20 degrees which cannot easily explain the X-ray modulation observed \cite{2006Scie...tmp}.
Moreover the collimation factor is 2-10 
times smaller than the maximum collimation reached in the mechanical beaming model \cite{2001ApJ...552L.109K}.
In case of isotropic emission, the super Eddington factor must be as high as $\sim100$ times the value obtainable 
in a standard accretion disc.
 These initial conditions also happen to be
highly unstable, as a slight variation in black hole mass and/or donor
mass makes the comparison with the observations unsatisfactory
(indicated with the crosses in fig.\,\ref{fig.1}).

If the initial period of binaries with a stellar mass black hole is increased around $3-4$days, the mass transfer rate become very violent 
in the beginning of the contact phase leading to rates of $\apgt 10^{-2}-10^{-4}\rm\,M_{\odot}\,yr^{-1}$. 
This is both a consequence of the very rapid expansion of the donor on the giant branch and of its mass which is often larger than the mass of the black hole.
The orbital period changes quickly surpassing the correct range in a few $\sim 10^{3}$yr, after which the mass of the black hole is larger than the donor
 and the evolution proceeds in a stable way with a period above 100 days.
Because the observed period is achieved for a very brief time during the evolution of these binaries, we consider them too unstable to reproduce the observations.

\begin{center}\begin{figure}
\centerline{\psfig{file=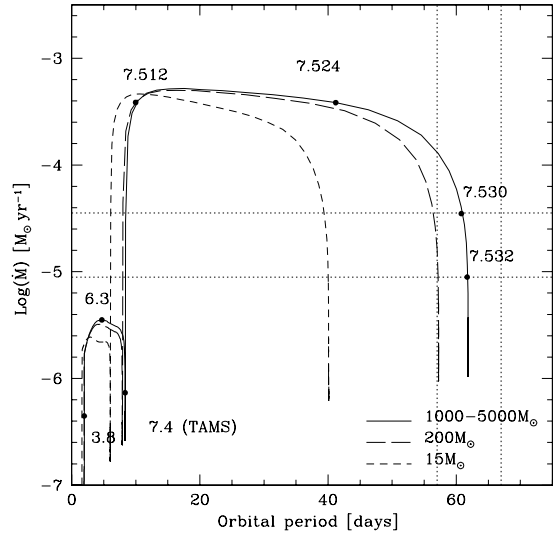,height=3in}} \caption{Evolution of the mass transfer rate as a function of the orbital period. The four small dashed straight lines creates a rectangle inside which the orbital period is within $2\sigma$ from the best value $62.0\pm 2.5$ days, and the mass transfer rate is in the correct range to produce bolometric luminosities in the range $5-20\times 10^{40}\ergs$. The three curves are respectively binaries with a donor star of $24\msun$ and a black hole of 15, 200 and 1000-5000$\msun$. The numbers in the plot give the age in Myr of the donor at several steps of the evolution assuming an IMBH mass of $\sim 1300\msun$. The small bump corresponds to a RLOF phase on the main sequence that lasts for 7.4 Myr until the donor reaches the terminal age main sequence (TAMS). The big bump is
produced after the main sequence and lasts only for $\sim 20000$yr. }\label{fig.2}
\end{figure} \end{center}

The largest area of the parameter space that satisfactorily reproduces
the observations are found for binaries with a donor of 22-25$\msun$
and a 200-5000$\msun\,$ IMBH. The initial orbital separation for these
binaries is 2-3 tidal radii, which is consistent with the tidal
capture of the donor by the IMBH (Hopman, Portegies Zwart \& Alexander 2004, Baumgardt et al. 2005).
The donor in these binaries start to transfer mass to the IMBH on
the main-sequence at the age of about 4\,Myr. By this time the rate of
mass transfer is still rather low ($\dot{M} \aplt 2 \times
10^{-6}\rm\,\msun/yr$), though sufficient to power a bright X-ray source
with $L_x \aplt 10^{40}$\,erg/s \cite{2004MNRAS.355..413P,2005MNRAS.364..344P}.
After tidal capture however, the main-sequence phase
lasts for 3-4\,Myr. In figure\,\ref{fig.2} we show the evolution of
the mass-transfer rate as a function of the orbital period.  
The binaries are initialized at short orbital period (in the lower left
corner of the figure) and evolve to higher mass transfer rate and
larger orbital period. The small feature in the lower left corner,
until an orbital period of about 10\,days, is the main-sequence
evolution. The broads shoulder is caused by the evolution of the donor
on the giant branch.

As soon as the donor ascends the giant branch the mass transfer rate
increases by about two orders of magnitude to $\dot{M} =
10^{-5}-10^{-3}\msun\rm\,yr^{-1}$, giving rise to a very bright
phase, with $L_x \sim 10^{41}-10^{42}$\,erg/s. This phase, however, only lasts
for 1-$2\times 10^4$\,years.  Near the end of this phase, the
rate of mass transfer drops as the orbital period increases.  At an
age of 7.2-9.0\,Myr the binaries reach an orbital period within
$2\sigma$ of the observed range of 57-67\,days and with a mass
transfer rate of $\dot{M} = 1$-$3\times 10^{-5}\rm\,\msun/yr$.  In
addition, the mean density of the donor in this time interval is $\rho =
3$-$5\times10^{-5}\rm\,g\,cm^{-3}$, also consistent with the
observations.

The binaries in this rather narrow window of initial conditions are
consistent with the observed age of the star cluster MGG-11, they
produce the observed luminosity of $L_{bol} = 5-20\times 10^{40}$\,erg/s
and their orbital period are within $2\sigma$ of the observed
periodicity. We therefore firmly conclude that the X-ray source M82
X-1 is satisfactorily explained by a binary in which a 22-25$\,\msun\,$
giant donor star transfers mass via Roche-lobe overflow to a
200-5000$\,\msun\,$ black hole. The lower limit of 200-400$\,\msun\,$ still needs a mild beaming or super Eddington isotropic emission (a factor ~2--8) to cope with the observations. 
The upper limit to the black hole mass
is ill constrained by the observed orbital period, because the orbital
period in a state of mass transfer is rather insensitive for black
holes of $M_{\rm bh} \apgt 1000\,\msun$.
If the initial orbital separation is wider ($4r_{t}<a_{i}<5r_{t}$) the RLOF phase begins near the terminal age main sequence, or when the star is already ascending the giant branch. In this situation the final period is much larger than observed

\section{Discussion}

The range for the donor mass in our best range of parameters
(22-25$\,\msun$) agrees well with recent $N$-body simulation \cite{2005astro.ph.11752B}. In these simulations the authors find that the typical mass of a
captured star is 25$\,\msun\,$ and that most tidal captures occur in the
first few million years after the formation of the IMBH.
We also note that even slightly larger initial separations $a>5r_t$ would result in significantly larger period than 62.0 days. A dynamical formation scenario of the binary such as dynamical friction or binary disruption \cite{2005astro.ph..8597B} is therefore unlikely.

A better age estimate for the star cluster MGG-11 could further
constrain our model. The observed range in cluster age is 7-12\,Myr
 \cite{2003ApJ...596..240M}. If the cluster happens to be on the young side
of this interval ($\aplt 9$\,Myr) we exclude the stellar mass
black hole as accreting object since our binary evolution models
systematically fails to reproduce the observed parameters.

The average time span over which a binary with an intermediate-mass
black hole that accretes from a stellar donor less massive than 30$\msun$
reaches luminosities in
excess of $L_x > 10^{40}$erg/s is a few times $10^4$\,years.  These
binaries spend about 4\,Myr as a relatively low luminosity x-ray
sources (with $L_x \simeq 10^{39}$ -- $10^{40}$\,erg/s). We would
therefore expect about 400 low luminosity sources on each bright ULX
(with $L_x > 10^{40}$\,erg/s).  An additional channel to make bright
sources but not the dim sources is the tidal capture for star more
massive than 30$\msun$. These sources spend the whole lifetime with
luminosities larger than $10^{40}\ergs$, increasing the expected
fraction of bright over dim ULXs. On the other hand a dynamical capture
during a close encounter can be another mechanism to form a dim ULX. This
process
has a $50\times$ larger cross section as tidal capture, and it may
be clear that a large fraction of the bright ULXs we observe are
formed by dynamical capture.  This is consistent with various ULX
catalogs \cite{2005A&A...429.1125L,2002ApJS..143...25C}, which
give 198 relatively dim ($L_x < 10^{40}\ergs$) versus 31 brighter sources.
Moreover the dynamical capture scenario will create also a large fraction
of detached systems with the donor under-filling its Roche lobe and
transferring mass on the black hole through the stellar wind. This will
produce luminosities between $10^{36}$ and $10^{39}\ergs$, in the range
of normal X-ray binaries.

With an average lifetime of $10^4$\,years and a present population of
$\sim 0.1$ per galaxy \cite{2004ApJS..154..519S} gives a formation
rate of about 1 per million years per galaxy.  These estimates are
based on all ULXs, which include the relatively dim $L_x \simeq
10^{39}$ -- $10^{40}\ergs$ x-ray sources.  We adopt a donor of $20\,\msun$,
which has a main-sequence lifetime of about 7\,Myr. The first 3\,Myr
of the star cluster is spend in forming an intermediate mass black
hole. The capture event is most likely to occur within about a Myr
after the formation of the IMBH \cite{2005astro.ph.11752B}, leaving
about 3--4\,Myr for the donor to transfer mass to the black hole.  The
formation rate of x-ray sources with $L_x > 10^{40}\ergs$ is then $R \simeq
0.3$ -- $1 \times 10^{-7}$ per year.

The average density of spiral galaxies, blue elliptical, starbursts
and irregular galaxies is about 6.1 per Mpc$^3$ (assuming $h = 0.76$)
\cite{1995AJ....110.2700V,1999ApJ...512L...9M}.  The observable
lifetime of a radio pulsar is about $10^7$ years
\cite{1985MNRAS.213..613L} which would allow the detection more than 5
to 17 binaries with an intermediate mass black hole and a observable
radio pulsar within the 1.2 Mpc range of LOFAR
\cite{2004...PhD...vanLeewuen}.  Finally we have to
correct for the beaming of the radio pulsar which, with a beaming
factor of 0.2 results in 1--3 potential detections with LOFAR.

The general consistency between the observations, the result of the
detailed $N$-body simulations and our binary evolution calculations
strongly constrains the nature of the ultra-luminous X-ray source in
M82 X-1 to a 22-25$\,\msun\,$ evolved donor star that transfers mass to
a 200-5000$\,\msun\,$ black hole.

\vskip 1.0 cm  We are
grateful to Phil Kaaret, Holger Baumgardt, Tom Maccarone, Monica Colpi, Luca Zampieri and  Tal Alexander for
stimulating discussions.  This work was supported by NOVA, NWO, KNAW,
IoA theory rolling grant from PPARC
\newcommand{\nat}{Nat}
\newcommand{\mnras}{MNRAS}
\newcommand{\aj}{AJ}
\newcommand{\pasp}{PASP}
\newcommand{\aap}{A\&A}
\newcommand{\apjl}{ApJ}
\newcommand{\apj}{ApJ}
\newcommand{\apjs}{ApJS}
\newcommand{\pasj}{PASJ}
\newcommand{\aaps}{AAPS}
\newcommand{\science}{Science}

\end{document}